\documentclass{elsart}

\usepackage[authoryear]{natbib}

\usepackage{graphics}
\usepackage{graphicx}
\usepackage{epsfig}

\usepackage{amssymb}

\begin{document}

\newcommand{\pam}{\textsf{PAMELA}}
\newcommand{\antip}{$\overline{p}$}
\newcommand{\posit}{$e^+$}

\begin{frontmatter}



\title{Launch of the Space experiment \pam .}


\author[a]{M.~Casolino\corauthref{cor}},
\corauth[cor]{Corresponding author}
\ead{Marco.Casolino@roma2.infn.it}
\author[a]{P.~Picozza},
\author[a]{F.~Altamura},
\author[a]{A.~Basili},
\author[a]{N.~De~Simone},
\author[a]{V.~Di~Felice},
\author[a]{M.P.~De~Pascale},
\author[a]{L.~Marcelli},
\author[a]{M.~Minori},
\author[a]{M.~Nagni},
\author[a]{R.~Sparvoli}
\author[b]{A.M.~Galper},
\author[b]{V.V.~Mikhailov},
\author[b]{M.F.~Runtso},
\author[b]{S.A.~Voronov},
\author[b]{Y.T.~Yurkin},
\author[b]{V.G.~Zverev}
\author[c]{G.~Castellini},
\author[d]{O.~Adriani},
\author[d]{L.~Bonechi},
\author[d]{M.~Bongi},
\author[d]{E.~Taddei},
\author[d]{E.~Vannuccini},
\author[d]{D.~Fedele},
\author[d]{P.~Papini},
\author[d]{S.B.~Ricciarini},
\author[d]{P.~Spillantini},
\author[e]{M.~Ambriola},
\author[e]{F.~Cafagna},
\author[e]{C.~De Marzo\corauthref{dec}}, \corauth[dec]{Deceased.}
\author[f]{G.C.~Barbarino},
\author[f]{D.~Campana},
\author[f]{G.~De Rosa},
\author[f]{G.~Osteria},
\author[f]{S.~Russo},
\author[g]{G.A.~Bazilevskaja},
\author[g]{A.N.~Kvashnin},
\author[g]{O.~Maksumov},
\author[g]{S.~Misin},
\author[g]{Yu.I.~Stozhkov},
\author[h]{E.A.~Bogomolov},
\author[h]{S.Yu.~Krutkov},
\author[h]{N.N.~Nikonov},
\author[i]{V.~Bonvicini},
\author[i]{M.~Boezio},
\author[i]{J.~Lundquist},
\author[i]{E.~Mocchiutti},
\author[i]{A.~Vacchi},
\author[i]{G.~Zampa},
\author[i]{N.~Zampa},
\author[j]{L.~Bongiorno},
\author[j]{M.~Ricci},
\author[k]{P.~Carlson},
\author[k]{P.~Hofverberg},
\author[k]{J.~Lund},
\author[k]{S.~Orsi},
\author[k]{M.~Pearce}
\author[l]{W.~Menn},
\author[l]{M.~Simon},
\address[a]{INFN, Structure of Rome ``Tor Vergata" and Physics
Department of University of Rome ``Tor Vergata", Via della Ricerca
Scientifica 1, I-00133 Rome, Italy}
\address[b]{Moscow Engineering and Physics Institute,
Kashirskoe Shosse 31, RU-115409 Moscow, Russia}
\address[c]{IFAC, Via Madonna del Piano 10, I-50019 Sesto Fiorentino, Florence, Italy}
\address[d]{INFN, Structure of
Florence and Physics Department of University of Florence, Via
Sansone 1, I-50019 Sesto Fiorentino, Florence, Italy}
\address[e]{INFN, Structure of Bari and Physics
Department of University of Bari, Via Amendola 173, I-70126 Bari,
Italy}
\address[f]{INFN, Structure of Naples and Physics Department of University of Naples ``Federico II",  Via Cintia, I-80126 Naples, Italy}
\address[g]{Lebedev Physical Institute, Leninsky Prospekt 53, RU-119991 Moscow, Russia}
\address[h]{Ioffe Physical Technical Institute,  Polytekhnicheskaya 26, RU-194021 St. Petersburg, Russia}
\address[i]{INFN, Structure of Trieste and Physics
Department of University of Trieste, Via A. Valerio 2, I-34127
Trieste, Italy}
\address[j]{INFN, Laboratori Nazionali di Frascati, Via Enrico Fermi 40,  I-00044 Frascati, Italy}
\address[k]{KTH, Department of Physics,
Albanova University Centre, SE-10691 Stockholm, Sweden}
\address[l]{Universit\"{a}t Siegen,  D-57068 Siegen, Germany}

  \begin{abstract}
 \pam\ is a satellite borne experiment designed to study with great accuracy
cosmic rays of galactic, solar, and trapped nature in a   wide energy range (protons: 80 MeV-700 GeV, electrons 50 MeV-400 GeV). Main objective is the study of
the antimatter component: antiprotons (80 MeV-190 GeV), positrons (50 MeV-270 GeV) and  search for antimatter with a precision of the order of $10^{-8}$). The experiment, housed
on board the  Russian Resurs-DK1 satellite,  was launched on June,
  $15$ 2006 in a $350\times600~km$ orbit with an inclination of
70 degrees.  The detector  is composed of  a series of
scintillator  counters arranged at the extremities of a permanent
magnet spectrometer to provide charge, Time-of-Flight and rigidity
information.   Lepton/hadron identification is performed  by a
Silicon-Tungsten calorimeter and a Neutron detector placed at the
bottom of the device. An Anticounter system is used offline to
reject false triggers coming from the satellite. In self-trigger
mode the Calorimeter, the neutron detector and a shower tail
catcher are  capable of an independent measure of the lepton
component up to 2 TeV.  In this work we describe the experiment,
 its scientific objectives and the performance in the first months after  launch.
\end{abstract}


Accepted for publication on Adv. Space Res.
http://dx.doi.org/10.1016/j.asr.2007.07.023

\end{frontmatter}

\section{Introduction}

\label{sec:intro} The Wizard collaboration is a scientific program
devoted to the study of cosmic rays through balloon and
satellite-borne devices. Aims involve the precise determination of
the antiproton \cite{boe97} and positron \cite{boe00} spectrum,
search of antimatter, measurement of low energy trapped and solar
cosmic rays with the NINA-1 \cite{nina} and NINA-2 \cite{nina2}
satellite experiments. Other research on board Mir and
International Space Station has involved the measurement of the
radiation environment,  the nuclear abundances and the
investigation of the Light Flash \cite{nat} phenomenon with the
Sileye experiments \cite{sil2,sil3}. \pam\ is the largest and most
complex device built insofar by the collaboration, with the
broadest scientific goals.  In this work we describe the
scientific objectives,  the detector  and the first in-flight
performance of \pam .

\section{Scientific Objectives}

\pam\ aims to measure in great detail the  cosmic ray component at 1 AU (Astronomical Unit).
Its $70^o$, $350\times 600 $ km orbit makes it  particularly suited to   study items of galactic, heliospheric and trapped nature.
 Indeed for its versatility and detector redundancy \pam\ is capable to address at the same time a number of different cosmic ray issues
 ranging over a very wide energy range, from the trapped particles in the Van Allen Belts, to electrons of Jovian origin, to the study of the
 antimatter component. Here we briefly describe the main scientific objectives of the experiment.

\subsection{Antimatter research.} 
\begin{figure}
\begin{center}
\epsfig{file=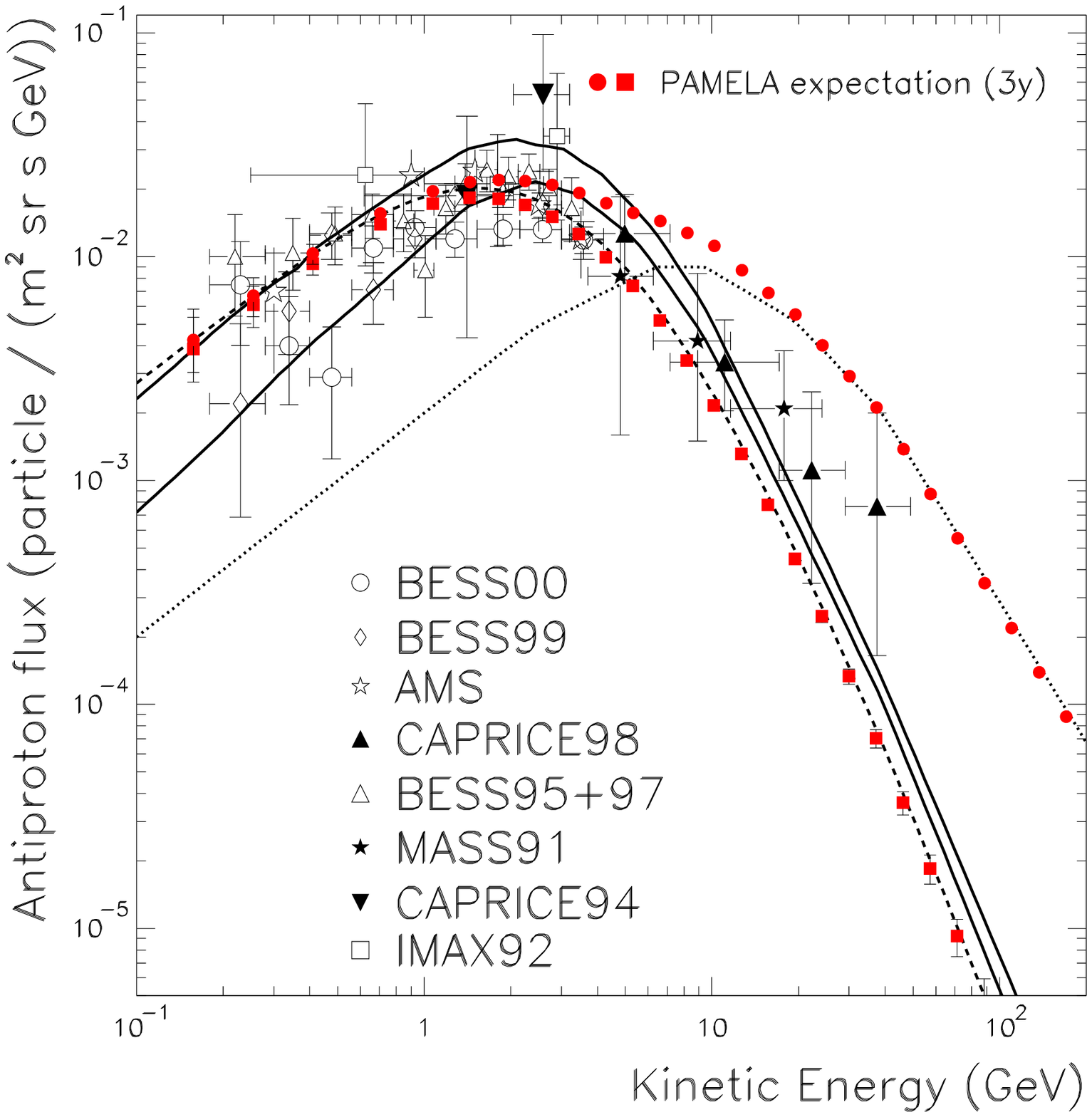, width=14.cm}
  \caption{ Recent experimental \antip\
spectra (BES\-S00 and BES\-S99 ~\cite{asa02}, AMS~\cite{agu02},
CAPRICE98~\cite{boe01a}, BES\-S95+97~\cite{ori00},
MASS91~\cite{bas99}, CAPRICE94~\cite{boe97}, IMAX92~\cite{mit96})
along with theoretical calculations for pure \antip\ secondary
production (solid lines: \cite{sim98}, dashed line: \cite{ber99b})
and for pure \antip\ primary production (dotted line:
\cite{ull99}, assuming the annihilation of neutralinos of mass
964~GeV/c$^2$). } \label{pbflu}
\end{center}
\end{figure}

\begin{figure}
\begin{center}
\epsfig{file=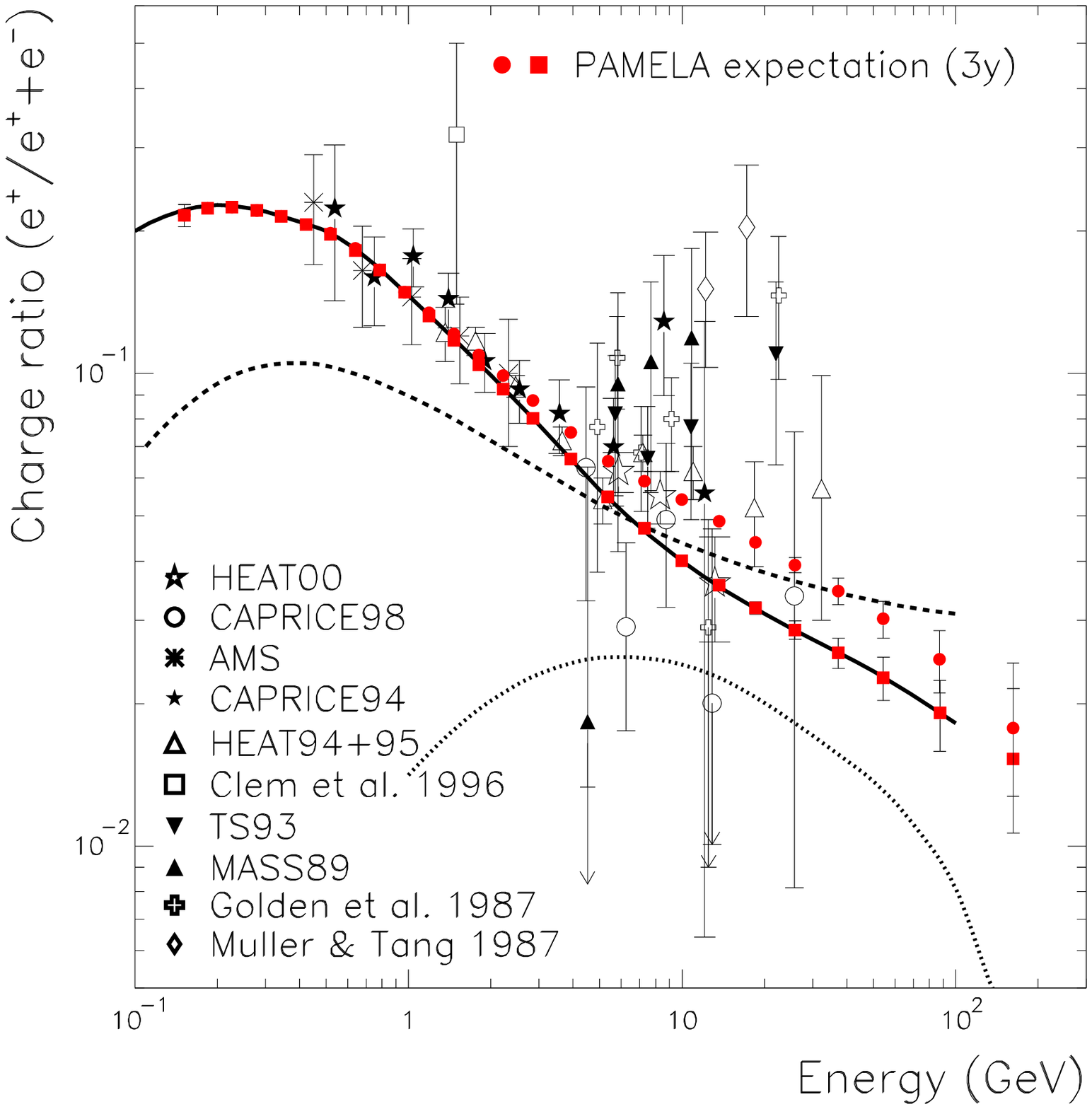, width=14.cm}
  \caption{    The positron fraction as a function
of energy measured by several experiments
(\cite{gol87,mul87,cle96} and MASS89~\cite{gol94},
TS93~\cite{gol96}, HEAT94+95~\cite{bar98}, CAPRICE94~\cite{boe00},
AMS~\cite{alc00}, CAPRICE98~\cite{boe01b}, HEAT00~\cite{bea04}).
The dashed \cite{pro82} and the solid \cite{mos98} lines are
calculations of the secondary positron fraction. The dotted line
is a possible contribution from annihilation of neutralinos of
mass 336~GeV/c$^2$ \cite{bal99}. The expected PAMELA performance,
in case of a pure secondary component (full boxes) and of an
additional primary component (full circles), are indicated in both
panels. Only statistical errors are included in the expected
PAMELA data.} \label{pbflu2}
\end{center}
\end{figure}

The study of the antiparticle component (\antip, \posit)   of cosmic rays  is the main scientific goal of \pam .  A long term and detailed study of the antiparticle spectrum over a very wide energy spectrum
 will allow to shed light over several questions of cosmic ray physics,  from particle production and propagation in the galaxy to
 charge dependent modulation in the heliosphere to dark matter detection.  In Figure \ref{pbflu} and \ref{pbflu2} are  shown the current status of
  the antiproton and positron measurements compared with \pam\ expected measurements in three years. In each case the two curves refer
   to a secondary only hypothesis with an additional contribution of a neutralino  annihilation. Also cosmological
    issues related to detection of a dark matter signature and search for antimatter (\pam\  will search for ${\overline{He}}$ with a
    sensitivity of $\approx 10^-8$) will therefore be addressed with this device.

\subsubsection{\bf Antiprotons} \pam\ detectable energy spectrum of \antip\ ranges from 80 MeV  to 190 GeV. Although the quality of \antip\ data has been improving in the recent years,  a   measurement of the energy spectrum
 of \antip\ will allow to greatly reduce the systematic error between the different balloon measurements, to study the
  phenomenon of charge dependent solar modulation,  and will for the first time explore the energy range beyond $\simeq 4 0$ GeV.
Possible excesses over the  expected secondary spectrum could be attributed to neutralino annihilation; \cite{profumo,ull99,donato}
  show that \pam\ is capable of detecting an excess of antiprotons due to neutralino annihilation in models  compatible with
   the WMAP measurements. Also \cite{lionetto} estimate that \pam\ will be able to detect a supersymmetric signal in many minimal supergravity  (mSUGRA)  models.
   The possibility to extract a neutralino annihilation signal from the background depends on the parameters used, the   boost
   factor  (BF) and the galactic proton spectrum.
\\  Charge dependent solar modulation, observed with the BESS balloon flights at Sun field reversal \cite{asa02} will be monitored
 during the period of recovery going from the $23^{rd}$ solar minimum going to the $24^{th}$ solar maximum.
Also the existence, intensity and stability of secondary antiproton belts \cite{miyasaka}, produced by the interaction of
 cosmic rays with the atmosphere will be measured.

\subsubsection{positrons}

A precise measurement of the positron energy spectrum is needed to distinguish dark matter annihilation from other galactic sources such as hadronic
 production in giant molecular clouds, $e^+/e^-$ production in nearby pulsars  or decay from radioactive nuclei produced in supernova explosions.  \pam\ is capable to detect   \posit\ in the energy range   50 MeV to 270 GeV.
Possibilities for dark matter detection in the positron channel
depend strongly on the nature of dark matter, its cross section
and the local inhomogeneity of the distribution. Hooper and Silk
\citetext{2004} perform different estimation of  \pam\ sensitivity
according to different hypothesis of the dark matter  component:
detection is possible  in case of an  higgsino of mass up to 220
GeV (with BF=1) and   to 380 GeV (with BF=5). Kaluza Klein models
\cite{hooper2} would give a  positron flux above secondary
production  increasing above 20 GeV and thus clearly compatible
with \pam\ observational parameters. In case of a bino-like
particle, as supposed by Minimal Supersymmetric Standard Model,
\pam\ is sensible to cross sections of the order of $2-3\times
10^{-26}$ (again, depending of BF). In case of Kaluza Klein
excitations of the Standard Model the sensitivity of \pam\ is for
particles up to 350 and 550 GeV. In the hypothesis of the littlest
Higgs model with T parity, the dark matter candidate is a heavy
photon which annihilates mainly into weak gauge bosons in turn
producing positrons. In \cite{asano} is  shown that \pam\ will be
able to identify this signal if the mass of the particle is below
120 GeV and the BF is 5.  Hisano et al.,  \citetext{2006} assume a
heavy wino-like dark matter component, detectable with \pam\ in
the positron spectrum (and with much more difficulty in the
antiproton channel) for mass of the wino above 300 GeV. This model
predicts that if the neutralino has a mass of 2 TeV the positron
flux increases by several orders of magnitude due to resonance of
the annihilation cross section in $W^+W^-$ and $ZZ$: in this
scenario not only such a signal would be visible by \pam\ but also
be consistent with the increase of positrons measured by HEAT
\cite{heat}.
 In conclusion a detailed measurement of the
 positron spectrum, its spectral features and its dependence from solar modulation will either provide evidence for a dark matter signature or
 strongly constrain and discard many existing models.

\subsection{\bf Galactic Cosmic Rays}
Proton and electron spectra will be measured  in detail with \pam
. Also light nuclei (up to O) are   detectable with the
scintillator system. In this way it is possible to study with high
statistics the secondary/primary cosmic ray nuclear and isotopic
abundances such as B/C,  Be/C,   Li/C and $^3He/ ^4He$. These
measurements will constrain existing production and propagation
models in the galaxy, providing detailed information on the
galactic structure and the various mechanisms involved.

\subsection{\bf Solar modulation of GCR}
Launch of \pam\ occurred   in the recovery  phase of solar minimum
toward cycle solar maximum of cycle 24. In this period it will be
possible to observe solar modulation of    galactic cosmic rays
due to increasing solar activity. A long term measurement of the
proton, electron and nuclear flux at 1 AU  will provide
information on  propagation phenomena occurring in the
heliosphere. As already mentioned, the possibility to identify the
antiparticle spectra will allow to study also charge dependent
solar modulation effects.

\subsection{\bf Trapped particles}

The   $70^o$ orbit of the Resurs-DK1 satellite allows for
continuous monitoring of the electron and proton belts. The high
energy ($>80 MeV$ component of the proton belt, crossed in the
South Atlantic region will be monitored in detail with the
magnetic spectrometer. Using the scintillator counting  rates   it
will be possible to   extend measurements of the particle spectra
to lower energies using the range method. Montecarlo simulations
have shown that the coincidence of the two layers of the topmost
scintillator (S1) allows \pam\ to detect $e^-$  from 3.5 MeV and
$p$ from 36 MeV. Coincidence between S1 and the central
scintillator (S2) allows us to measure   integral spectra of  9.5
$e^-$ and 63 MeV p. In this way it will be possible to  perform a
detailed mapping of the Van Allen Belts showing spectral and
geometrical  features. Also the neutron component will be
measured, although some care needs to be taken to estimate the
background coming from proton interaction with the main body of
the satellite.

\subsection{\bf Solar energetic particles}

We expect about 10 significant
solar events during the experiment's lifetime with energy high enough to be detectable \cite{shea}. The rate of
background particles hitting the top trigger scintillator (S1)
could be very high for intense solar events, hence a different
trigger configuration has to be set in  these cases. The usual
trigger involves a   coincidence of S1 with those  located before
(S2) and after (S3) the tracker. During solar particle events a
devoted trigger mask (e.g. using only S2 and S3)  can  be
programmed from ground\footnote{This can occur using information
coming from the satellite monitoring system (e.g. SOHO, ACE,
GOES). In this way observation and memory filling would therefore
vary according to the event type (impulsive, gradual) and
intensity.}. The observation of solar energetic particle (SEP)
events with a magnetic spectrometer will allow several aspects of
solar and heliospheric cosmic ray physics to be  addressed for the
first time.

\subsubsection{Electrons and Positrons}
\label{sec:positron} Positrons are produced mainly in the decay of
$\pi^{+}$ coming from nuclear reactions occurring at the flare
site. Up to now, they have only been measured indirectly by remote
sensing of the gamma ray annihilation line at 511~keV. Using the
magnetic spectrometer of \pam\ it will be possible to separately
analyze the high energy tail of the electron and positron spectra
at 1 Astronomical Unit (AU) obtaining information both on particle
production and charge dependent propagation in the heliosphere in perturbed conditions
of  Solar Particle Events.

\subsubsection{Protons}
\label{sec:proton} \pam\ will be able to measure the spectrum of
cosmic-ray protons from 80~MeV up to almost 1~TeV and  therefore
will be able to measure the solar component over  a very wide
energy range (where the upper limit will be limited  by
statistics).
These measurements will be correlated with other
instruments placed in different points of the Earth's
magnetosphere to give information on the acceleration and
propagation mechanisms of SEP events. Up to now there has been no
direct measurement~\cite{miroshnichenko} of the high energy
($>$1~GeV) proton component of SEPs. The importance of a direct
measurement of this spectrum is related to the fact~\cite{ryan}
that there are many solar events where the energy of protons is
above the highest ($\simeq$100 MeV) detectable energy range of
current spacecrafts,  but is below the detection threshold of
ground Neutron Monitors~\cite{bazilevskaya}. However, over the
\pam\ energy range, it will be possible to examine the turnover of
the spectrum, where we find the limit of acceleration processes at
the Sun. The instrument has a maximum trigger rate of about 60~Hz
and a geometrical factor of 21.5 cm$^{2}$ sr. This implies that we
will be able to read all events with an integral flux (above
80~MeV) up to 4 particles/$cm^{2} s\: sr$. For such events we
expect about 2$\times$10$^{6}$ particles/day (assuming a spectral
index of $\gamma$ = 3 we have 2$\times$10$^{3}$ events / day above
1~GeV).

\subsubsection{Nuclei}
\label{sec:nuclear}  \pam\  can identify light nuclei up to Carbon
and isotopes of Hydrogen and Helium. Thus we can investigate the
light nuclear component related to SEP events over a wide energy
range. This should contribute to  establish whether there are
differences in the  composition of the high energy (1 GeV) ions
  to the low energy component ($\simeq$ 20 MeV)   producing $\gamma $ rays or the quiescent solar corona\cite{ryan05}.
\\ Applying the same estimates as above, we can expect
$\simeq$10$^{4}$ $^{4}$He and $\simeq$10$^{2}$ $^{3}$He nuclei for
gradual events, and more for impulsive (often $^3He$ enriched) ones. Such a high
statistics will allow us to examine in detail the amount of the
$^{3}$He and deuterium (up to 3 GeV/c). These measurements will
help us to better understand the selective acceleration processes
in the higher energy impulsive~\cite{reames} events.

\subsubsection{Neutrons}
\label{sec:neutron} Neutrons are produced in nuclear reactions at
the flare site and can reach the Earth before decaying \cite{chupp}. Although
there is no devoted trigger for neutrons in \pam, the background
counting of the neutron detector will measure in great detail the
temporal profile and distribution of solar neutrons. The
background counting system keeps track of the  number of
neutrons which hit the neutron detector in the time elapsed since
last trigger. The counter is   reset each time it is read allowing
for a precise measurement of background neutron conditions during
the mission. On the occurrence of solar events, neutrons are
expected to reach Earth before protons as they have no charge.
They are not deflected by any magnetic field and will be directly
recorded by \pam\ (if it is not in Earth's shadow).

\subsubsection{Lowering of the geomagnetic cutoff}
\label{sec:geo} The high inclination of the orbit of the
Resurs-DK1 satellite will allow \pam\ to study \cite{ogliore,
leske} the variations of cosmic ray geomagnetic cutoff due to the
interaction of the SEP events with the geomagnetic field.

\subsection{\bf Jovian Electrons}
\label{sec:jovian} Since the discovery made by the Pioneer 10
satellite of Jovian electrons at about 1~AU from
Jupiter~\cite{simpson, eraker}, with an energy between 1 and
25~MeV, several interplanetary missions have measured this
component of cosmic rays. Currently we know that Jupiter is the
strongest electron source at low energies (below 25~MeV) in the
heliosphere within a radius of 11~AUs. Its spectrum has a power
law with spectral index $\gamma$ = 1.65, increasing above 25~MeV,
where the galactic component becomes dominant. At 1~AU from the
Sun the IMP-8 satellite could detect Jovian electrons in the range
between 0.6 and 16~MeV and measure their modulation by the passage
of Coronal Interaction Regions (CIR) with 27 days
periodicity~\cite{eraker, chenette}. There are also long term
modulation effects related to the Earth-Jupiter synodic year of 13
months duration. In fact, since Jovian electrons follow the
interplanetary magnetic field lines, when the two planets are on
the same solar wind spiral line, the electron transit from Jupiter
to the Earth is eased and the flux increases. On the other side,
when the two planets lie on different spiral lines the electron
flux decreases. For \pam\ the minimum threshold energy for
electron detection is 50~MeV. In this  energy range, however,
geomagnetic shielding will reduce the active observation time
reducing total counts. Nevertheless it will be possible to study
for the first time the high energy Jovian electron component and
test the hypothesis of reacceleration at the solar wind
Termination Shock (TS). It is known that  cosmic rays originating
outside the heliosphere can be accelerated at the solar wind TS.
This applies also to Jovian electrons, which are transported
outward by the solar wind, reach the TS and   undergo shock
acceleration thus increasing their energy. Some of these electrons
are scattered back in the heliosphere. The position of the shock
(still unknown and placed at about 80--100~AU) can affect the
reaccelerated electron spectrum~\cite{potgieter,pot2}. Overall,
Jovian electrons are dominant in the energy range 50-70 MeV and
decrease in intensity to  about 1$\%$ of the total galactic flux
above 70 MeV. This component  can however be extracted from the
galactic background with  observation periods of the order of
four-five months \cite{Casolino}. In addition it is possible that
the reacceleration of electrons at the solar wind TS is modulated
by the solar cycle;
  three years of observations in the recovering phase of   the solar minimum
  should show such effects.

\subsection{\bf High energy lepton component}
The calorimeter can provide an independent trigger to \pam\ for
high energy releases due to showers occurring in it: a signal  is
generated  with the release of energy above 150 mip in  all the 24
views of planes from 7 to 18. With this requirement the
geometrical factor of the calorimeter self-trigger is 400 $cm^2
sr$  if events  coming from the satellite are rejected. In this
way it is possible to study the    electron and positron flux in
the energy range between 300 GeV  and 2 TeV, where measurements
are currently scarce \cite{koba}. At this energy discrimination
with hadrons is performed with topological and energetic
discrimination of the shower development in the calorimeter
coupled with neutron information coming from the neutron detector.
This is because   neutron  production cross-section  in an e.m.
cascade is lower than in a hadronic cascade\cite{galper}.

\section{Instrument Description}

\begin{figure}

\begin{center}
\epsfig{file=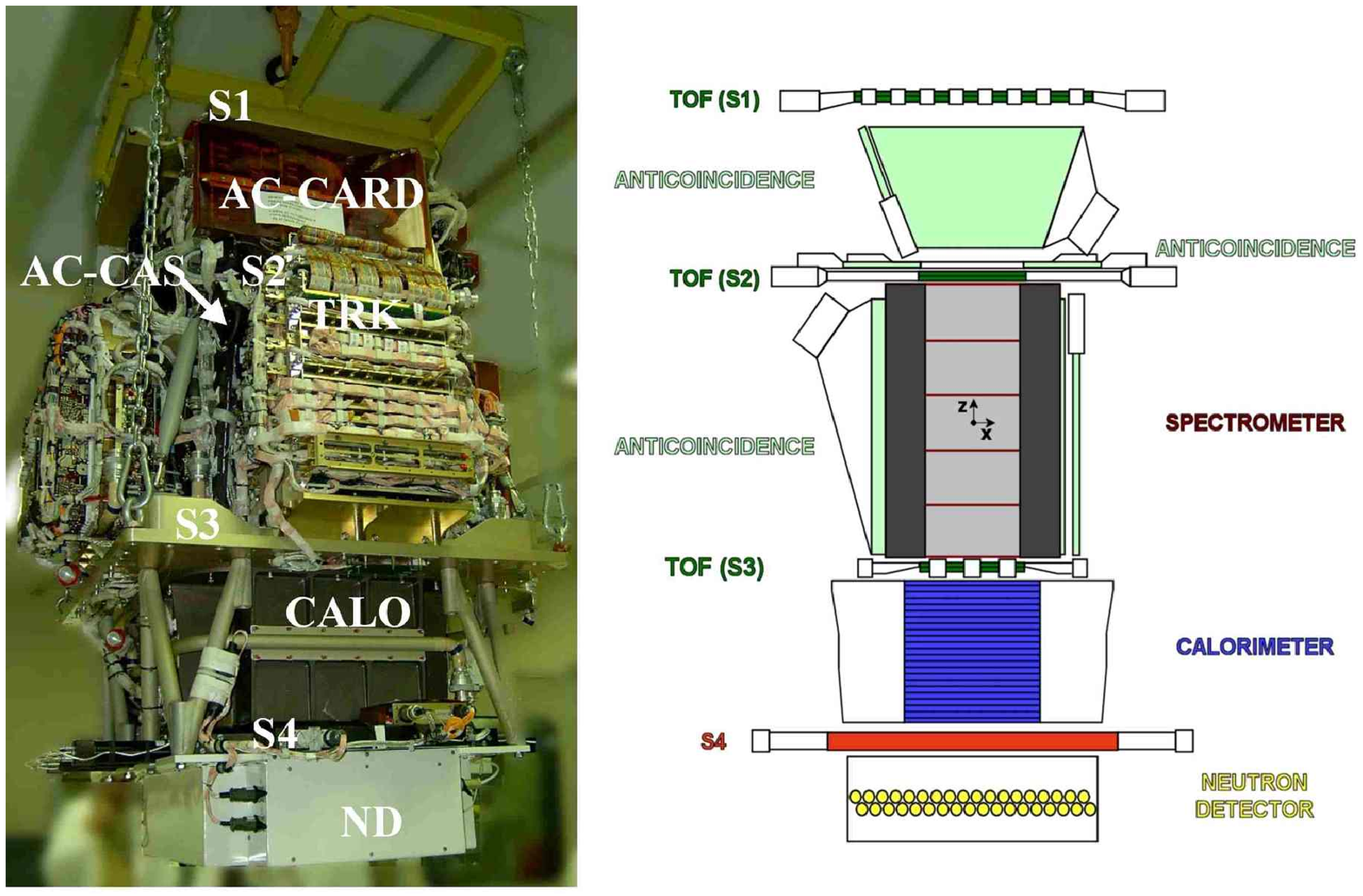, width=14.cm}
 \caption{Left: Photo of
the  \pam\ detector during the final  integration phase in Tor
Vergata clean room facilities, Rome. It is possible to discern,
from top to bottom,  the topmost scintillator system, S1,  the
electronic crates around the magnet spectrometer, the baseplate
(to which \pam\ is suspended by chains), the black structure
housing the Si-W calorimeter, S4 tail scintillator and the neutron
detector.  Right: scheme - approximately to scale with the picture
-  of  the detectors composing \pam. } \label{scheme2}
\end{center}
\end{figure}

\begin{figure}
\begin{center}
\epsfig{file=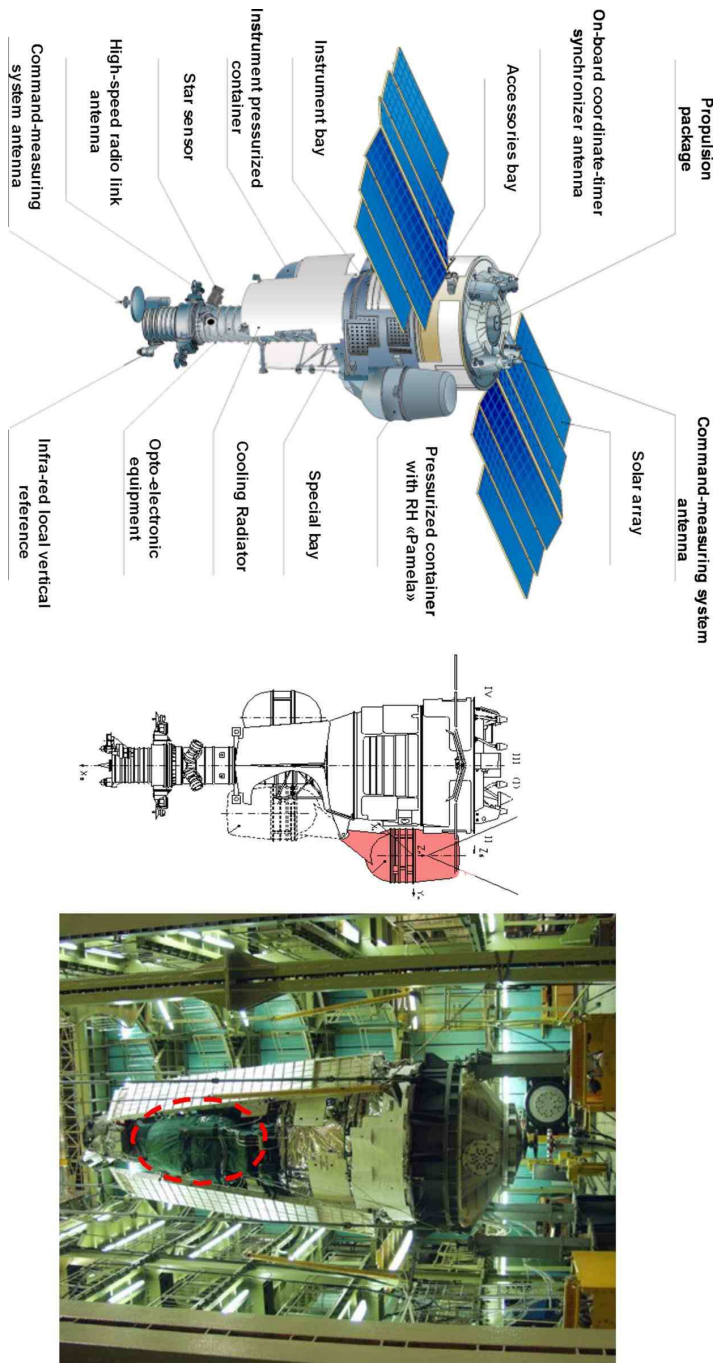, width=18.cm,  angle=-90}
  \caption{Left: Scheme of
the Resurs-DK1 satellite.   \pam\ is located in the pressurized
container on the right of the picture. In the center panel it is
possible to see the container in the launch position and in the
extended (cosmic ray acquisition) configuration. In the right
panel it is possible to see a picture of the satellite in the
assembly facility  in Samara. The picture   is rotated 180 degrees
to compare the photo with the scheme. The dashed circle  shows the
location of \pam\ pressurized container in the launch position.}
\label{scheme}
\end{center}
\end{figure}

\begin{figure}
\begin{center}
\epsfig{file=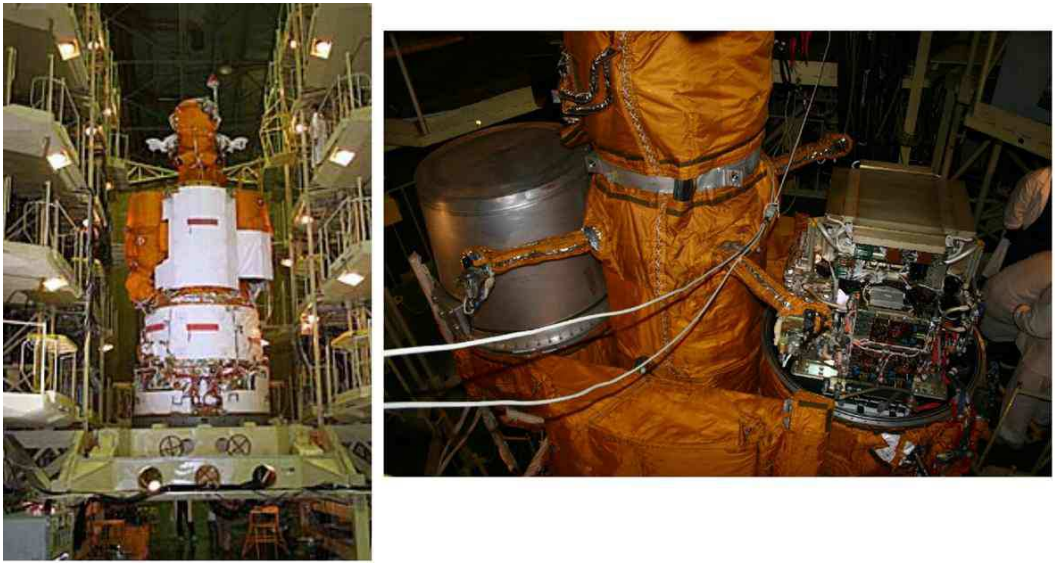, width=14.cm}
  \caption{Left: Photo of
Resurs in the final integration phase in Baikonur. It is possible
to discern the   the optical sensor on top,  the two pressurized
containers on the sides, and the white heat cooling panel in the
forefront. Right: close up picture of the integration phase of
\pam\ in the pressurized container (right in picture).  }
\label{scheme1}
\end{center}
\end{figure}

In this section we  describe   the main characteristics of \pam\
detector; a more detailed description of the device and the data
handling  can be found in \cite{picozza, cpu, yoda}. The device
(Figure \ref{scheme2}) is constituted by a number of highly
redundant detectors capable of identifying particles providing
charge, mass,  rigidity and beta    over a very wide energy range.
The instrument is built around a  permanent magnet  with a silicon
microstrip   tracker  with  a scintillator system to provide
trigger, charge and time of flight information. A silicon-tungsten
calorimeter is used to perform hadron/lepton separation. A shower
tail catcher and a neutron detector at the bottom of the apparatus
increase this separation. An anticounter system is used to reject
spurious events in the off-line phase. Around the detectors are
housed the readout electronics, the interfaces with the CPU and
all primary and secondary power supplies. All systems (power
supply, readout boards etc.)  are redundant with the exception of
the CPU which is more tolerant to failures. The system is enclosed
in a pressurized container (Figure \ref{scheme},\ref{scheme1})
located on one side of the Resurs-DK satellite. In a twin
pressurized  container is  housed the   Arina experiment, devoted
to the study of the low energy trapped electron and proton
component. Total weight of \pam\ is 470 kg;  power consumption is
355 W, geometrical factor is 21.5$cm^2 sr$.

\subsection{Scintillator / Time of Flight}
The scintillator system\cite{tof} provides trigger for the
particles and time of flight information for incoming particles.
There are three scintillators layers, each composed by two
orthogonal   planes  divided in various bars (8 for S11, 6 for
S12, 2 for S21 and S12 and 3 for S32 and S33) for a total of 6
planes  and 48 phototubes (each bar is read by two phototubes). S1
and S3 bars  are 7 mm thick and  S2 bars are 5 mm. Interplanar
distance between  S1-S3 of  77.3 cm  results in  a TOF
determination of 250 ps precision for protons and 70 ps for C
nuclei (determined with beam tests in GSI), allowing separation of
electrons from antiprotons up to $\simeq 1$ GeV and albedo
rejection. The scintillator system is also capable of providing
charge information up to $Z=8$.
\subsection{Magnetic Spectrometer}
The permanent magnet \cite{track} is composed of 5 blocks, each
divided in 12 segments of  Nd-Fe-B alloy with a residual
magnetization of 1.3 T arranged to provide an almost uniform
magnetic field along the $y$ direction. The size of the cavity is
$13.1\times 16.1\times 44.5\: cm^3$, with a mean magnetic field of
0.43 T. Six layers of $300\mu \: m$ thick double-sided microstrip
silicon detectors are used to measure particle deflection with
$3.0\pm 0.1\: \mu m$ and $11.5\pm 0.6 \: \mu m$ precision in the
bending and non-bending views. Each layer is made  by three
ladders, each composed by two  $5.33\times 7.00\: cm^2$ sensors
coupled to a VA1 front-end hybrid circuit. Maximum Detectable
Rigidity (MDR) was measured on CERN proton beam to be $\simeq 1\:
TV$.

\subsection{Silicon Tungsten Calorimeter}
Lepton/Hadron discrimination is performed by the Silicon Tungsten
sampling calorimeter \cite{calo} located on the bottom of \pam\ .
It is composed of 44 silicon layers  interleaved by 22  0.26 cm
thick Tungsten plates. Each silicon layer is composed arranging
$3\times 3$ wafers, each of $80\times 80\times  .380\: mm^3$ and
segmented in 32 strips, for a total of 96 strips/plane. 22 planes
are used for  the X view and 22  for  the Y view in order to
provide topological and energetic information of the shower
development in the calorimeter. Tungsten was chosen in order to
maximize electromagnetic radiation  lengths (16.3 $X_o$)
minimizing hadronic interaction length (0.6 $\lambda $). The
CR1.4P ASIC chip is used for front end electronics, providing a
dynamic range of 1400 mips (minimum ionizing particles) and
allowing nuclear identification up to Iron.
\subsection{Shower tail scintillator}
This  scintillator ($1\times 48\times 48\: cm^3$)   is located
below the calorimeter and is used to improve hadron/lepton
discrimination by measuring the energy not contained in the shower
of the calorimeter. It can also function as a standalone trigger
for the neutron detector.
\subsection{Neutron Detector}
The $60 \times  55 \times 15\: cm^3$ neutron  detector is composed
by 36 $^3He$ tubes arranged in two layers and surrounded by
polyethylene shielding and a 'U' shaped cadmium layer to remove
thermal neutrons not coming from the calorimeter. It is used to
improve lepton/hadron identification by detecting the number of
neutrons produced in the hadronic and electromagnetic cascades.
Since the former have a much higher neutron cross section than the
latter, where neutron production comes essentially through nuclear
photofission, it is estimated that \pam\ overall identification
capability is improved by a factor 10. As already mentioned, the
neutron detector is   used  to measure neutron field in Low Earth
Orbit and in case of solar particle events as well as in the high
energy lepton measurement.
\subsection{Anticoincidence System}
To reject  spurious triggers due to interaction with the main body of the satellite, \pam\ is shielded by a number of scintillators
used with anticoincidence functions\cite{anti,pearce}.  CARD anticoincidence system is composed of four 8 mm thick scintillators
located in the area between S1 and S2. CAT scintillator is placed on top of the magnet: it is composed by a single piece with a central
 hole where the magnet cavity is located and  read out by 8 phototubes. Four scintillators,  arranged on the sides of the magnet, make the CAS
 lateral anticoincidence system.
\section{Integration and Launch}
Pamela was integrated in INFN - Tor Vergata clean room facilities, Rome; tests involved first each subsystem separately and subsequently  the
 whole apparatus simulating all interactions with the satellite using an Electronic Ground Support Equipment. Final tests involved cosmic ray
  acquisitions with muons for a total of about 480 hours.
The device was then   shipped to Tskb Progress plant, in Samara
(Russia), for installation in a pressurized container  on board the
Resurs-DK satellite for final tests. Also in this case acquisitions with cosmic muons (140 hours)
 have been performed and have shown the correct functioning of the apparatus, which was then integrated with the pressurized
  container and the satellite. In 2006 the final integration phase took place in Baikonur cosmodrome in Kazakstan.

 Launch occurred   with      Soyuz-U  rocket  on June $15^{th}$ 2006,
  08:00:00.193 UTC, orbital insertion was in a parking orbit of semimajor axis of 6642 km.
   Final boost occurred on June $18^{th}$ 2006 when the orbit was raised to a semimajor axis of 6828 km.

Altitude of the  satellite is usually minimum  in the northern
hemisphere and maximum  in the southern hemisphere since its
primary goal is to perform earth observations and   resolution of
the pictures increases at lower altitudes. To compensate for
atmospheric drag, the altitude of the satellite is   periodically
reboosted by vernier engines. To perform this manoeuvre the
pressurized container housing \pam\ is folded back in the launch
position, the satellite is rotated $180^o$ on its longitudinal
axis and then engines are started. Reboost frequency depends from
orbital decay, due to atmospheric drag, in turn caused   by solar
activity. Up to October 2006 the activity has been low with only
two  small Solar Particle Events, so there has not been the need
to perform this maneuver yet.  Resurs-DK1  oscillates on its
longitudinal axis when performing Earth observations:   a detailed
information of the attitude of the satellite is provided to the
CPU of \pam\  in order to know the orientation of the detector
with precision of $\simeq 1$ degree.

\section{In flight data}
\begin{figure}
\epsfig{file=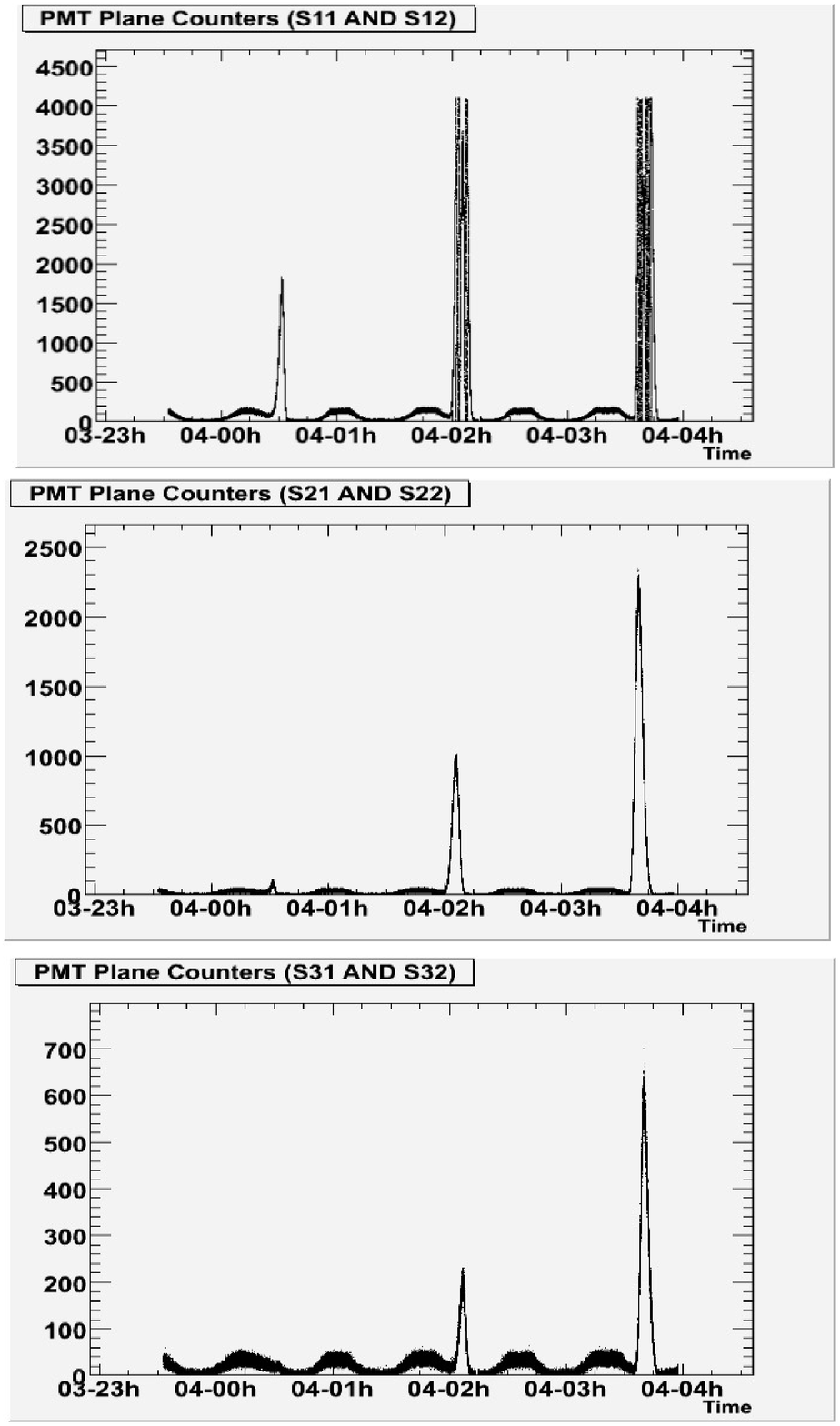, width=14.cm}

\begin{center} \caption{Counting rates of \pam\
(ADC channels) as a function of time.} \label{trigrate}
\end{center}
\end{figure}

\begin{figure}
\epsfig{file=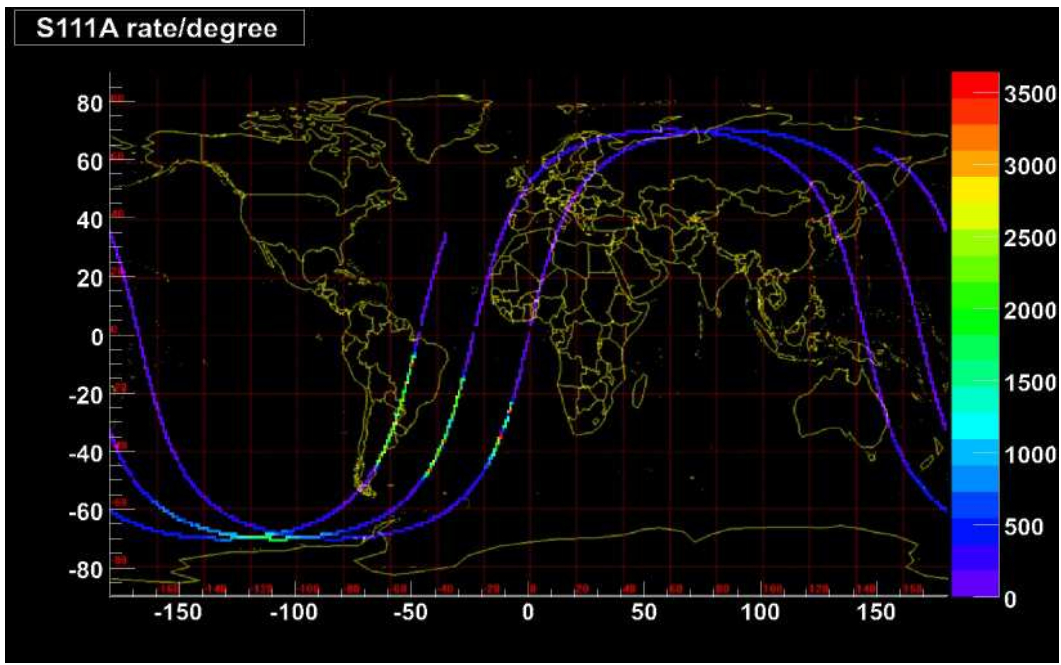, width=14.cm}

\begin{center}
\caption{Ground track of \pam\ with counting rate of S11*S12
trigger. Note the particle rate increase in the South Atlantic
Anomaly} \label{groundtrack}
\end{center}
\end{figure}

\begin{figure}
\epsfig{file=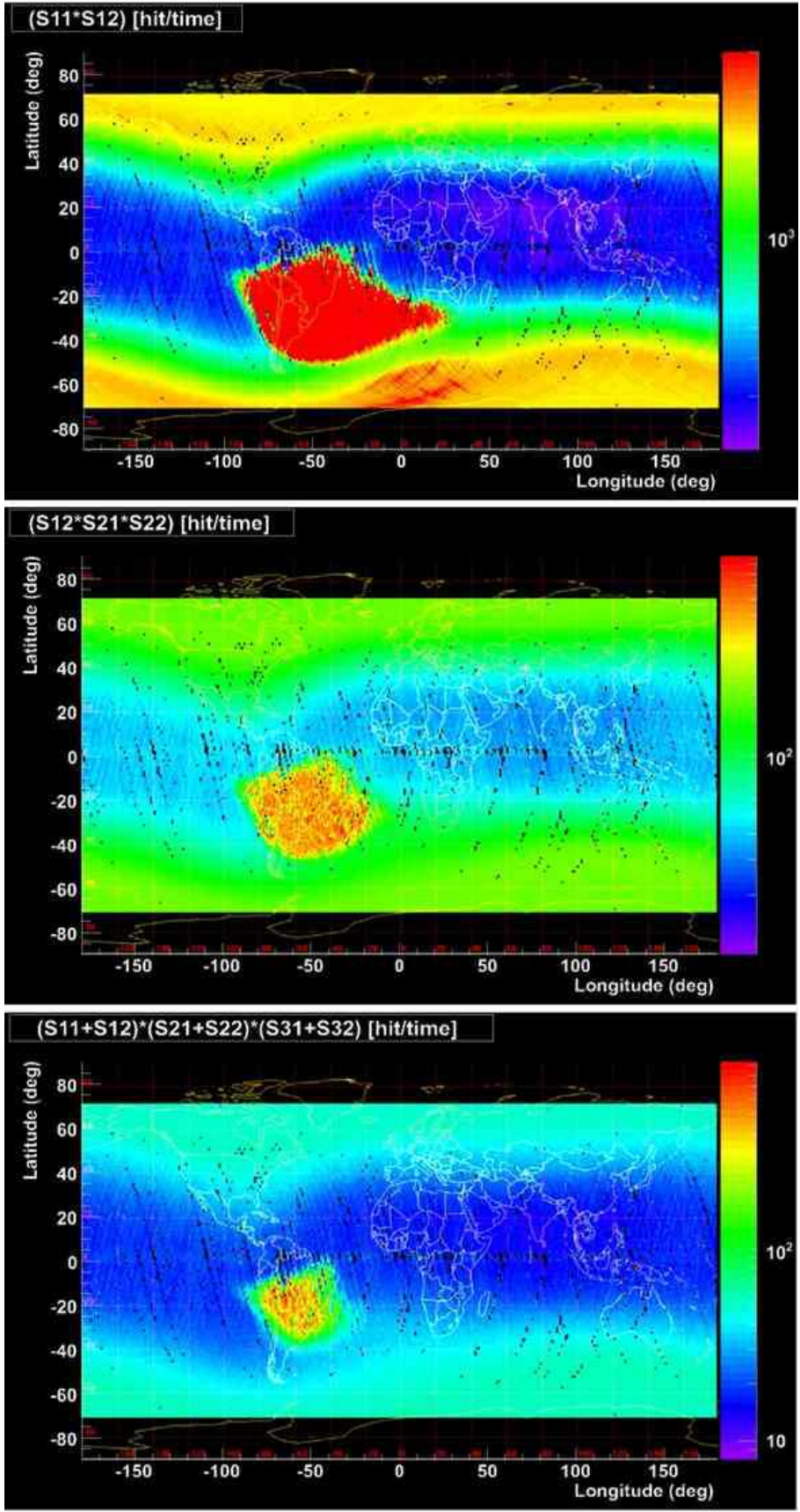, width=12.cm}

\begin{center}
\caption{Counting rates of various trigger counters of the
scintillators. Top Panel: S11*S12 (p 36 MeV, $e^-$ 2.5 MeV).
Center Panel: S12*21*S22 (p  63 MeV, $e^-$ 9.5MeV). Bottom panel:
(S11*S12)*(S21+S22)*(S31+S32) (p 80 MeV, $e^-$ 50 MeV). }
\label{mappe}
\end{center}
\end{figure}

\pam\ was first switched on  June, $26^{th}$ 2006.
A typical behaviour of the  acquisition of the device is shown in Figure
 \ref{trigrate}. The three panels show the counting rate of the three planes
  and  correspond  to particles of increasing energy: S11*S12 is triggered by  36
   MeV protons and  2.5 MeV $e^-$,   S21*S22 requires protons and electrons of 9.5 and 63 MeV respectively
   and S31*S32 requires protons and electrons  of 80 MeV and 50 MeV (lower energy particles may penetrate the
   detector from the sides and increase the trigger rate). The higher energy cut is evident in the counting rate of the
   scintillators: the first SAA passage, corresponding to the easternmost passage is absent in the S3 counting rate. The other two
    passages in the SAA saturate several times the ADC counting rate of the S1 scintillator but not the other two scintillators. This is
    consistent with the power law spectrum of trapped particles in the SAA and the    geometrical  ratio between the two scintillators.
Also S2 and S3 counting rate ratio in the anomaly is about 5,
consistent with a particle flux ratio evaluated with AP8 algorithm
\cite{ae8}.  Outside the SAA it is possible to see the increase of
particle rate at the geomagnetic poles due to the lower
geomagnetic cutoff. The highest rates are found when the satellite
crosses the trapped components of the Van Allen Belts. In Figure
\ref{groundtrack} it is possible to see the same trigger rate as a
function of the ground track  of \pam . It  is possible to see the
inclination of the orbit of  satellite and the particle rate
increase in the polar and South Atlantic regions.   The pause in
the acquisition at the equator are  due to the calibration of the
subdetectors usually performed   during the ascending phase to
avoid crossing the radiation belts. In Figure \ref{mappe} is shown
\pam\ world particle rate for S1*S2 counter, exhibiting   the high
latitude electron radiation belts and the proton belt in the South
Atlantic Anomaly. Note that the northern electron belt is less
intense than the southern one because \pam\ orbit is $\simeq 350$
km in that region as compared to the $\simeq 600$ km in the south.
In the same Figure, panel 2 is shown the same map for
$S12*S21*S22$ trigger configuration, corresponding to 9.5 MeV
electrons and 63 MeV protons. It is possibile to see how the
intensity of the electron belts is reduced and size of the SAA is
smaller as expected for higher energy protons.

\section{Conclusions}
\pam\ was successfully launched on June 2006 and is  currently
operational in Low Earth Orbit. The satellite and the detectors
are   functioning correctly. It it expected that data from \pam\
will provide information on several items of cosmic ray physics,
from antimatter to solar and trapped particles.

\newpage

\end{document}